\documentclass[aps,prl,twocolumn,showpacs,superscriptaddress,groupedaddress]{revtex4}  
\usepackage{graphicx}  
\usepackage{dcolumn}   
\usepackage{bm}        
\usepackage{amssymb}   
\usepackage[dvipsnames,usenames]{color}
\usepackage{color}

\begin{document}


\newcommand{\etal}{{\sl et al.}}
\newcommand{\ie}{{\sl i.e.}}
\newcommand{\sto}{SrTiO$_3$} 
\newcommand{\lto}{LaTiO$_3$}
\newcommand{\lao}{LaAlO$_3$}
\newcommand{\lno}{LaNiO$_3$}
\newcommand{\nith}{Ni$^{3+}$}
\newcommand{\nitw}{Ni$^{2+}$}
\newcommand{\otw}{O$^{2-}$}
\newcommand{\alo}{AlO$ _2 $}
\newcommand{\tio}{TiO$ _2 $}
\newcommand{\eg}{$e_{g}$}
\newcommand{\tg}{$t_{2g}$}
\newcommand{\dzt}{$d_{z^2}$}
\newcommand{\dxtyt}{$d_{x^2-y^2}$}
\newcommand{\dxy}{$d_{xy}$}
\newcommand{\dxz}{$d_{xz}$}
\newcommand{\dyz}{$d_{yz}$}
\newcommand{\egp}{$e_{g}'$}
\newcommand{\ag}{$a_{1g}$}
\newcommand{\mub}{$\mu_{\rm B}$}
\newcommand{\ef}{$E_{\rm F}$}
\newcommand{\alao}{$a_{\rm LAO}$}
\newcommand{\asto}{$a_{\rm STO}$}
\newcommand{\nst}{$N_{\rm STO}$}
\newcommand{\lnnlam}{(LNO)$_N$/(LAO)$_M$}
\newcommand{\lxtlaf}{(La$X$O$_3$)$_2$/(LaAlO$_3$)$_4$}

\title{Design of Mott and topological phases on buckled $3d$-oxide honeycomb lattices}

\author{David Doennig}
\affiliation{Forschungs-Neutronenquelle Heinz Maier-Leibnitz (FRM II), Technische Universit\"at M\"unchen, Lichtenbergstra\ss{}e 1, 85748 Garching, Germany}
\author{Santu Baidya}
\affiliation{Department of Physics and Center for Nanointegration Duisburg-Essen (CENIDE), University of Duisburg-Essen, Lotharstr. 1, 47057 Duisburg, Germany}
\author{Warren E. Pickett}
\affiliation{Department of Physics, University of California Davis, One Shields Avenue, Davis, CA 95616, U.S.A.}
\author{Rossitza Pentcheva}
\email{Rossitza.Pentcheva@uni-due.de}
\affiliation{Department of Physics and Center for Nanointegration Duisburg-Essen (CENIDE), University of Duisburg-Essen, Lotharstr. 1, 47057 Duisburg, Germany}
\affiliation{Forschungs-Neutronenquelle Heinz Maier-Leibnitz (FRM II), Technische Universit\"at M\"unchen, Lichtenbergstra\ss{}e 1, 85748 Garching, Germany}
\date{\today}

\begin{abstract}
Perovskite bilayers with (111)-orientation combine a honeycomb lattice as a key feature with the strongly correlated, multiorbital nature of electrons in transition metal oxides. In a systematic DFT+$U$ study of (111)-oriented (La$X$O$_3$)$_2$/(LaAlO$_3$)$_4$ superlattices, we
establish trends in the evolution of ground states versus band filling in
(111)-oriented (La$X$O$_3$)$_2$/(LaAlO$_3$)$_4$ superlattices, with $X$ spanning the 
entire $3d$ transition metal series. The competition between local quasi-cubic and global triangular symmetry triggers unanticipated broken symmetry phases, with mechanisms ranging from Jahn-Teller 
distortions, to charge-, spin-, and orbital-ordering. LaMnO$_3$, where spin-orbit coupling opens a sizable  gap in the Dirac-point Fermi surface, emerges as a topological Chern insulator.

\end{abstract}

\maketitle

\section{Introduction}

Synthesis and characterization of atomically abrupt transition metal oxide (TMO) heterostructures 
have revealed a broad platform of unanticipated functionalities with potential to enhance next generation electronics 
and spintronics capabilities by controlling charge, spin, orbital, and lattice degrees of freedom at the 
nanoscale.\cite{hwang2012,Mannhart}  

The earlier focus on perovskite materials with the  (001) growth orientation - prominent examples here being e.g. 
the interface between LaAlO$_3$ and SrTiO$_3$ (LAO/STO)\cite{jpcm2010,zubko} 
or nickelate superlattices \cite{nickelates} 
-  has recently been extended to the (111) orientation.\cite{Middey2012,Gibert2012} In the latter, two triangular BO$_6$ 
sublattices form a  buckled honeycomb lattice, topologically equivalent to  graphene.  The honeycomb lattice itself introduces 
exotic possibilities: considering an alternating, next nearest neighbor (nnn) {\it imaginary} hopping amplitude, Haldane 
obtained a quantum spin Hall (QSH) system without explicit external field.\cite{Haldane88}  In the Kane-Mele elucidation\cite{KaneMele} 
of the honeycomb lattice, topological behavior originated  from
spin-orbit coupling (SOC) entanglement of band character. Later, Raghu {\it et al.}\cite{Raghu2008} proposed that topological
character can be generated by strong interactions (even in the mean field
approximation).  
Wright generalized Haldane's model to a {\it buckled honeycomb lattice} in which magnetic flux, alternating in orientation in 
neighboring cells, leads to a Chern (quantum anomalous Hall [QAH]) insulator\cite{Wright2013}  with topologically protected 
gapless edge states. These specific models are, however, challenging to realize in real materials, and viable Chern insulators
remain to be realized.

Compared to graphene and common topological insulators (TI), TMOs possess not only larger band gaps,  but offer an enormously 
richer palette of possibilities due to several distinctive features: correlated electron behavior causing spin, charge, and 
orbital instabilities, multi-orbital configurations combined with relativistic effects, viz. spin-orbit coupling (SOC). Below 
we demonstrate that the interplay of strong interactions and SOC effects produce specific spin-fermion systems as candidates for QSH or QAH systems in the $3d$ (111) bilayers with buckled honeycomb lattice.

The idea of constructing a (111) bilayer from perovskite TMO was introduced by Xiao {\it et al.} for $4d$ and $5d$ systems, 
pointing to possibilities for ``interface engineering of quantum anomalous Hall effects".\cite{Xiao2011} 
Building on this foundation, Yang {\it et al.}\cite{Yang2011} and R\"uegg and Fiete\cite{Ruegg2011} applied a tight-binding (TB) model for 
(LaNiO$_3$)$_2$/LAO, and demonstrated that in certain
ranges of parameters and magnetic order  topological insulating phases can
result from ordering of a complex combination of $e_g$ orbitals.\cite{Ruegg2011} 

Complementary to tight-binding models, material-specific density functional theory (DFT) can both contribute towards  a fundamental understanding as well as guide the search for actual materials realizations. 
Besides examples for $5d$ systems\cite{Lado2013}, 
DFT studies including strong local interaction effects (see below) have recently 
predicted a Dirac-point Fermi surface for  STO(111)\cite{Doennig2013} and  LNO(111) bilayers, quantum confined within 
LAO.\cite{Yang2011,Ruegg2011,Doennig2014} In these cases the Dirac point is `protected' by symmetry; sublattice symmetry breaking leads to gap-opening charge-disproportionated states.\cite{Doennig2013,Doennig2014}

To identify further promising systems as well as to elucidate
the underlying {\it design principles} of functionalities, we have 
explored systematically the effect of band filling on the electronic ground state in 
(111)-oriented \lxtlaf\ superlattices, where $X$ spans the range of trivalent $3d$ ions 
Ti-Cu.
Despite the fact that these systems, unlike LAO/STO(111), have nonpolar interfaces - i.e. there is no valence mismatch
across the interface, so all transition metal cations retain $X^{3+}$ configurations - unexpected phases proliferate. 
Competition between local pseudocubic symmetry and global trigonal symmetry 
as well as additional flexibility, provided by the magnetic and spin degrees 
of freedom of $3d$ ions, lead to a broad array of distinctive broken symmetry ground states, offering a platform to design 2D electronic functionalities.  

Moreover, while Chern insulators have so far been sought mainly in magnetically doped TIs\cite{Chang2013}, or very recently, by combining trivial magnetic insulators with a material with large SOC, either in double pervoskites\cite{Cook2014} or rock-salt compounds\cite{Garrity2014,Zhang2014}, two of our systems display simultaneous time reversal symmetry breaking and SOC-driven gap opening. These are first examples in a solely $3d$ system with gap sizes large enough to support room-temperature applications.  

\begin{figure*}[h!tbp]
\includegraphics[scale=0.70]{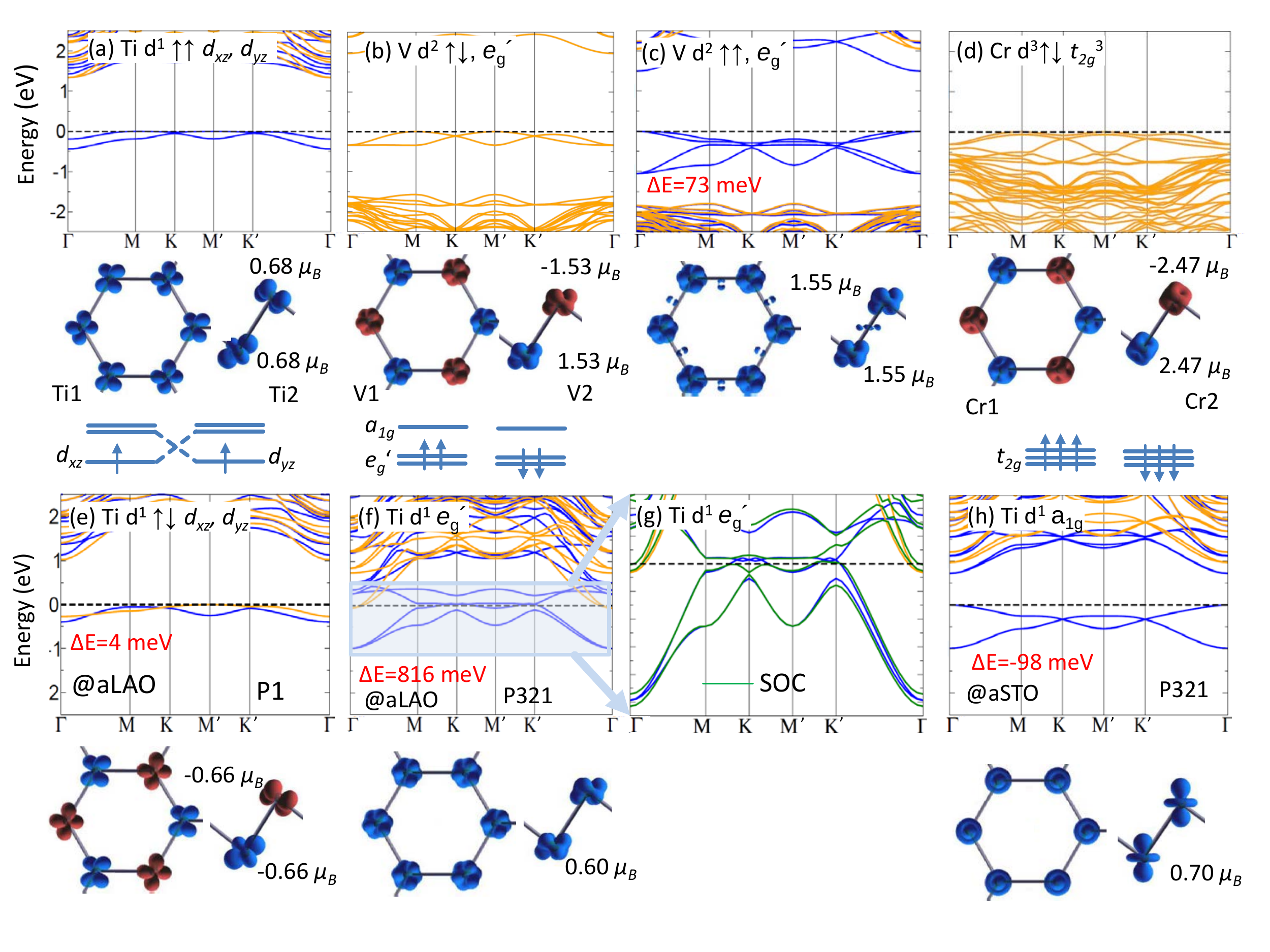}
\caption{Electronic ground (a-d) and selected metastable (e-h) states in \lxtlaf(111)
for \tg\ systems $X$=Ti, V, Cr. Shown are the band structures with
blue/orange denoting the majority/minority bands, and isosurfaces of
the spin density, with majority in blue and minority in red. In
cases (a,e,h) the integration range is \ef-1eV,\ef\ to emphasize the orbital polarization. Energies of metastable
states are provided in red. 
} 
\label{fig:t2g}
\end{figure*}

\begin{figure*}[h!tbp]
\includegraphics[scale=0.67]{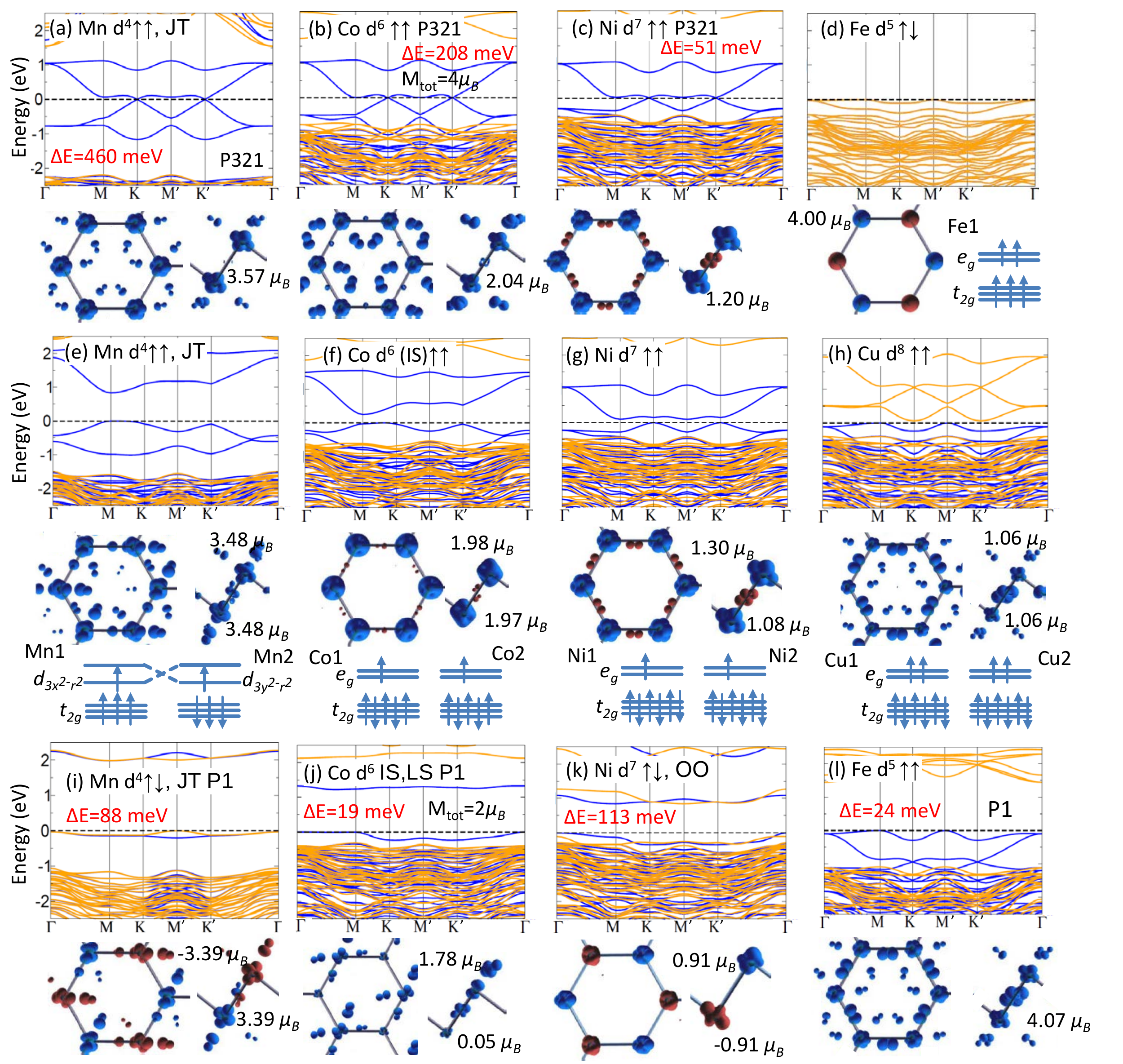}
\caption{ 
Presentation as in Fig. \ref{fig:t2g}, but for \eg\ systems $X$=Mn, Fe, Co, Ni,
and Cu.  A striking feature is the similarity in band structure of $X$=Mn, Co, Ni, despite the formally different band filling:  a Dirac point Fermi surface within P321 (a)-(c) and a gap opening due to symmetry breaking. The ground state results are presented in
(d)-(h).  
}
\label{fig:eg}
\end{figure*}

DFT calculations were performed on  (111)-oriented \lxtlaf\ superlattices with $X=3d$, using the all-electron full-potential linearized augmented-plane-wave (LAPW) method, as implemented in the WIEN2k code~\cite{wien2k}. For the exchange-correlation functional we used the generalized gradient approximation (GGA)~\cite{pbe96}. Static local electronic correlations were included in the GGA+$U$ approach~\cite{anisimov93} with $U=$5\,eV, $J=$0.7\,eV (for all $X=$ Ti-Cu 3$d$) and $U=8$\,eV (La 4$f$).  Systematic investigations of the influence of $U$ show that the results are robust with respect to variation of the $U$ parameter in a reasonable range of values. Additional calculations with the modified Becke-Johnson~\cite{mBJ} potential support the obtained electronic behavior.

The lateral lattice constant is fixed to \alao$=$3.79\,\AA, corresponding to superlattices grown on a LAO(111) substrate, unless otherwise stated. Octahedral tilts and distortions were fully taken into account when relaxing atomic positions, whether constrained to P321 symmetry or fully released to P1 symmetry. Additionally, the out-of-plane lattice parameter $c$ was optimized for all superlattices. Spin-orbit coupling (SOC) was treated using the second-variational method. AHC was calculated using wannier90\cite{wannier90,wannierAHC} interfaced with Wien2k\cite{wien2wannier}.

{\bf Trends across the $3d$ series:}
A central aspect in the \lxtlaf(111) honeycomb bilayers is their strong deviation  from their bulk analogs.  
Certain recurring features tied to the $t_{2g}$ and $e_g$ distinction can be identified, thus we discuss them separately. 

{\bf \tg\ systems:}
For the $t_{2g}$ subshell the dominating feature is a competition between local pseudo-cubic symmetry and the underlying 
threefold+inversion symmetry (``P321'') of the ideal bilayer.
The (111) bilayering 
reduces the octahedral symmetry to trigonal and splits the $t_{2g}$ orbital triplet into $a_{1g}$ + $e_g'$, with the former having zero angular 
momentum around the $\hat c$ bilayer axis while the latter orbital doublet forms a representation
for $m_{\ell} = \pm 1$ angular momentum. Breaking
this symmetry allows occupation of the cubic $d_{xy}$, $d_{yz}$, or  $d_{xz}$ orbitals.

$X$=Ti$^{3+}$ $3d^1$. The above scenario arises most vividly for the LaTiO$_3$ bilayer, 
which displays the richest behavior among the \tg\ systems.  
The ground state at the lateral lattice constant of LaAlO$_3$ (\alao), pictured in Fig.~\ref{fig:t2g}a),  is a ferromagnetic (FM) orbitally ordered
Mott insulator, displaying staggered $d_{xz}$,  $d_{yz}$ occupation and a very narrow (0.2 eV bandwidth) lower Hubbard band. This is in contrast to bulk LaTiO$_3$  which is a distorted $Pbnm$, G-type antiferromagnetic (AFM)  Mott insulator with  $1/\sqrt{3}(d_{xy}+d_{yz}+d_{zx})$ orbital order~\cite{Torrance1993,pavarini2004}. 
Consistent with this extremely localized character, the corresponding AFM state with the 
same orbital polarization, shown in Fig.~\ref{fig:t2g}e, is only 4 meV/u.c. higher in energy, suggesting
a weak exchange coupling of $\sim$1 meV.  

Constraining the symmetry to P321, thereby keeping the two Ti ions
related by symmetry, results in \egp\ orbital polarization with bands touching  at K and K$^{\prime}$ (Fig.~\ref{fig:t2g}f). This 
degeneracy protects the system against the Mott insulating
gap, presenting a special case of strong correlation effects being 
rendered ineffectual by an imposed symmetry. Despite its high energy cost (0.4 eV/Ti), this state is intriguing due to the unusual  direction reversal of bands in the vicinity of  K and K$^{\prime}$ points (note zoom-in of the band structure in Fig.~\ref{fig:t2g}g). This intertwining of bands suggests topological character. 
	
Inclusion of SOC with out-of-plane magnetization 
leads to a band inversion and a resulting gap  (green line in Fig.~\ref{fig:t2g}g) and a nonzero anomalous Hall conductivity (AHC).  
The disparity at K and K$^{\prime}$ signals the loss of
equivalence of the two Ti ions, reflected in 
 an surprisingly large orbital moment of  one of the ions: 0.11$\mu_B$
versus just 0.01$\mu_B$ on the other. 

Ti orbital polarization is highly susceptible to strain-tuning: applying tensile strain by imposing the lateral 
lattice constant of \sto\ tips the 
pseudo-cubic/trigonal symmetry balance, stabilizing occupation of the $a_{1g}$ orbital (Fig.~\ref{fig:t2g}h). 
The band structure just below the gap is comprised of two filled bands with Dirac
crossings at K and K$^{\prime}$, similar to the analogous LAO/STO(111) 
case,\cite{Doennig2014} where the  
$3d^{0.5}$ band filling fixes the Dirac points at \ef\ instead.

$X=$V$^{3+}$ $3d^2$.   
The AFM ground state of the LaVO$_3$-bilayer, displayed in Fig.~\ref{fig:t2g}b, is gapped due to occupation of the majority \egp\ doublet. This is insensitive to strain and at variance with the bulk G-type $d_{xz}$, $d_{yz}$ orbital ordering ($d_{xy}$ is occupied on all sites)\cite{pavarini2004,Raychaudhury2007}. Thus in this LaVO$_3$ bilayer
trigonal symmetry splitting dominates over the pseudocubic crystal field. 
The FM state (Fig.~\ref{fig:t2g}c) with the same orbital polarization is 73 meV/u.c. 
higher in energy, and has 
four bands topologically similar to those of the metastable 2LaTiO$_3$ case (Fig.~\ref{fig:t2g}f), 
with the difference that now the entire set of bands 
is filled.

$X=$Cr$^{3+}$ $3d^3$.
The Cr bilayer is electronically trivial: a half-filled \tg-band ($t_{2g,\uparrow}^3$, S=$\frac{3}{2}$), thus no orbital degrees of freedom, and antiferromagnetic order (Fig.~\ref{fig:t2g}d).  

{\bf \eg\ systems:}
Now we turn to the \eg\ systems.
Consistent with the tight-binding model of Xiao {\it et al.}\cite{Xiao2011} 
a distinctive set of four bands emerges for ferromagnetic coupling with an open $e_g$ subshell:
nearly flat bottom and top bands interconnected by two dispersive bands, providing a Dirac point 
crossing at the K and K$^{\prime}$ points and quadratic contact with the flat bands at the 
$\Gamma$ point (cf. Fig.~\ref{fig:eg}a-c,h,l). A key finding is that a pinning of the Dirac 
point at \ef\ is not solely determined by
band filling, but also by an interplay of orbital and spin degrees of freedom, as proven for the cases of $X=$Mn, Co and Ni. Equivalence of the two sublattices again becomes crucial.
This symmetry is found to be broken in {\sl all}  \eg\ system ground states where the Dirac point is initially
at the Fermi level. We identify distinct origins of symmetry breaking and the resulting gap opening in 
each system, as discussed below.

$X$=Mn$^{3+}$ $3d^4$. 
The LaMnO$_3$ bilayer renders one of the promising cases where, within P321 symmetry, the system 
exhibits a Dirac point Fermi surface within the $e_g$ bands (Fig.\ref{fig:eg}a). 
The high-spin (HS) Mn ion ($t_{2g,\uparrow}^3 e_{g,\uparrow}^1$ with the majority \eg-band  half-filled putting the Dirac point at \ef) is unstable 
to Jahn-Teller distortion. Releasing structural symmetry restrictions leads to an elongation of the apical Mn-O bond lengths to 
2.07-2.11 \AA\ and variation of the basal distances between 1.89-1.98 \AA, associated with
alternating $d_{3y^2-r^2}$,  $d_{3x^2-r ^2}$ occupation on the A and B sublattices (Fig. \ref{fig:eg}e). 
This symmetry breaking opens a gap of 0.8 eV and also lifts the quadratic band touching degeneracy at $\Gamma$. 
The Jahn-Teller distortion is also present in the AFM order (Fig. \ref{fig:eg}i), which is 88 meV/u.c. higher in energy.  
 The significantly flatter bands reflect electronic decoupling of the two sublattices, similar to the AFM LaNiO$_3$ bilayer (Fig. \ref{fig:eg}k), discussed below.

$X$=Fe$^{3+}$ $3d^5$. 
The ground state of the LaFeO$_3$ bilayer is a HS AFM band insulator with 
nearly spherically symmetric charge
and spin density on the Fe site, characteristic of a half-filled $3d$ band 
($t_{2g,\uparrow}^3 e_{g,\uparrow}^2$ with S=$\frac{5}{2}$, cf. Fig \ref{fig:eg}d). For comparison, bulk LaFeO$_3$ is a G-type AFM with orthorhombic $Pnma$ structure. The metastable FM 
configuration exhibits the previously discussed set of four bands, 
albeit now these are fully occupied for the majority spin channel (cf. Fig \ref{fig:eg}l).

$X$=Co$^{3+}$ $3d^6$. Bulk LaCoO$_3$ has a low spin (LS) ($t_{2g}^6$) narrow gap insulating ground state, but becomes ferromagnetic e.g. as a strained film \cite{cobalt}. A rich set of (metastable) states with respect to spin degrees of freedom can be anticipated and is indeed realized  in the LaCoO$_3$ bilayer.
Constraining symmetry to 
P321 renders another case
where the Fermi level is pinned at a Dirac point (Fig. \ref{fig:eg}b).  This state lies 0.21 eV above
the broken symmetry ground state of 
FM intermediate spin (IS) ($t_{2g}^5$, $e_{g}^1$) insulator with a moment of 1.97$\mu_B$ (Fig. \ref{fig:eg}f). 
Orbital ordering of the $e_g$ 
electron and $t_{2g}$ hole drives the breaking of symmetry, analogous to the case of Mn but with a smaller band gap.

A further metastable state, only 19 meV less favorable, exhibits a new type of {\sl spin state symmetry breaking} in
which the two Co sublattices assume IS and LS states with very flat bands accompanied by (and presumably caused by) a $d_{x^2-y^2}$ 
orbital occupation on the IS Co sublattice (Fig. \ref{fig:eg}j).

$X$=Ni$^{3+}$ $3d^7$. Bulk LaNiO$_3$ is a $R\bar{3}c$ correlated metal. \cite{Torrance1993}. 
Within P321 symmetry a Dirac-point Fermi surface is obtained for the 
LaNiO$_3$ bilayer (cf. Fig. \ref{fig:eg}c), 
as previously reported \cite{Yang2011,Ruegg2011,Doennig2014}. However, breaking the equivalency of the two triangular 
sublattices opens a gap of 0.25 eV at the Fermi level 
(cf. Fig. \ref{fig:eg}g)\cite{Doennig2014}. 
Here the mechanism is disproportionation of the Ni sublattice, expressed in different magnetic moments of 1.30 and 1.08 \mub. AFM coupling of the two bilayers results in flat bands (Fig. \ref{fig:eg}k), defining a band gap of $\sim 1$ eV with orbital polarization at the Ni sites, as recently observed in a NdNiO$_3$ bilayer.\cite{Middey2012} This illustrates how antiferromagnetic order provides the necessary decoupling of the two trigonal bilayers, analogous to the La$_2$NiAlO$_6$ double perovskite where the single triangular Ni-layers are separated by Al-layers \cite{Doennig2014}. 

$X$=Cu$^{3+}$ $3d^8$. This case yields a straightforward $e_{g,\uparrow}^2$ 
S=1 ion at half filling of the \eg\ bands, 
with large optical (spin-conserving) gap but with very low energy 
spin-flip excitations (Fig.~\ref{fig:eg}h, see also \cite{Xiao2011}).  Although Cu$^{3+}$ is
uncommon (bulk LaCuO$_3$ is metallic and must be
synthesized under pressure),\cite{karppinen} it might be stabilized by non-equilibrium epitaxial synthesis. 

Despite the formal difference in band filling for $X$=Mn, 
Co, and Ni, an unexpected analogy occurs in the 
electronic structure of these bilayers: a Dirac-point Fermi 
surface when constrained to P321 symmetry reverts to 
a gapped Mott insulator state, being driven by symmetry breaking 
 of distinct origin: a Jahn-Teller distortion in LaMnO$_3$; 
a spin transition to IS accompanied by orbital order in LaCoO$_3$; and a
charge disproportionation in LaNiO$_3$.

\begin{figure}[tbp]
\includegraphics[scale=0.60]{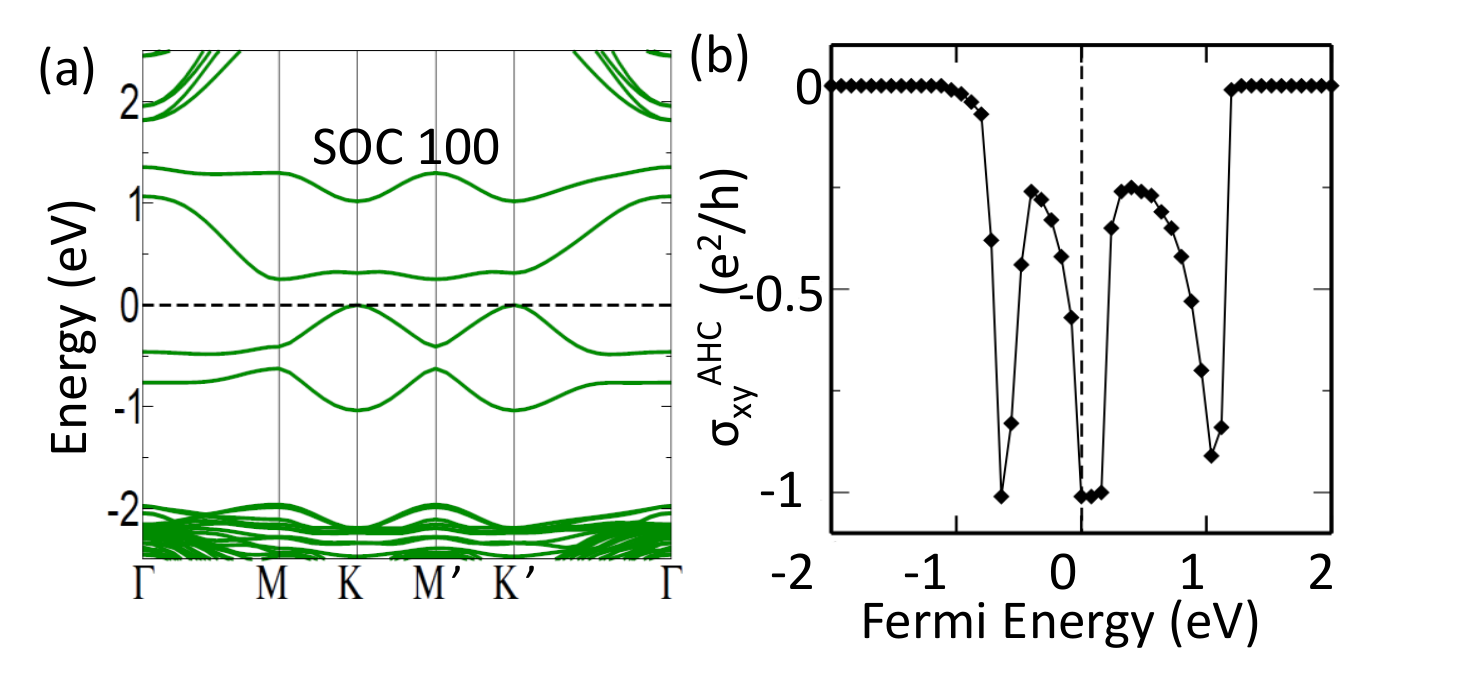}
\caption{Effect of SOC on (LaMnO$_3$)$_2$/(LaAlO$_3$)$_4$ a) band structure for magnetization along [100]; b) calculated AHC for a range of Fermi levels, plotted in units of (e$^{2}$/h), shows a Chern number of unity. }
\label{fig:MnSOC}
\end{figure}

{\it Spin-orbit coupling}.
Generally, no significant SOC is expected in the $3d$ series beyond minor 
band splittings and small magnetocrystalline anisotropy. However, an unusually strong effect arises in 
LaMnO$_3$ and LaCoO$_3$: SOC gaps both the Dirac points at K and K$^{\prime}$ as well as the quadratic band
touching at $\Gamma$. 
(LaMnO$_3$)$_2$ shows a stronger influence of SOC, with an in-plane 
[100] easy axis (Fig. \ref{fig:MnSOC}a), 
while (LaCoO$_3$)$_2$ has an out-of plane easy axis. 
The anomalous Hall conductivity\cite{wien2wannier,wannier90,wannierAHC} plotted versus (rigid band) 
chemical potential in (Fig. \ref{fig:MnSOC}b), determines (LaMnO$_3$)$_2$ as a
C=-1 Chern insulator.   The gap of 200 meV indicates the possibility of room temperature applications.

Apart from three cases of filled subshells that stabilize symmetric phases,
the buckled $3d$ honeycomb bilayers encounter and succumb to a variety of symmetry-breaking
forces: (1) orbital ordering including competition between occupation of real pseudocubic
orbitals ($d_{xy}$, $d_{xz}$, $d_{yz}$ for $t_{2g}$; $d_{x^2-y^2}$ and $d_{z^2}$ for $e_g$) 
and their triangular lattice counterparts; (2) charge disproportionation, (3) spin-state
differentiation, and (4) Jahn-Teller distortion. 

Spin-orbit coupling, typically important rather in $4d$ and $5d$ than $3d$ systems, is found to gap the Dirac point Fermi surface,
in two members of this series $X$=Mn and Co, inducing  
Chern insulating phases with gap size (for LaMnO$_3$) allowing consideration for room temperature applications. 

This study outlines the variety of possibilities for unusual ground states and topological 
behavior in this ``$3d$ palette" of two-dimensional oxide honeycomb lattices.  The predicted trends should stimulate 
research on the experimental realization of these phases, which can be combined into
numerous hybrid systems using multilayer growth techniques. Doping of these materials
will provide further possibilities to tune functionalities, interesting for basic research reasons as well as
for potential applications.

\begin{acknowledgments}
Discussions with Binghai Yan are gratefully acknowledged. This work was partially supported by the German Science Foundation within SFB/TR80, project G3
(to R.P. and D.D.) and by U.S. Department of Energy Grant No. DE-FG02-04ER46111 (to W.E.P.).
\end{acknowledgments}


\begin{thebibliography}{99}

\bibitem{hwang2012}
  H. Y. Hwang, Y. Iwasa, M. Kawasaki, B. Keimer, N. Nagaosa and Y. Tokura,  Nat. Mater. {\bf 11}, 103 (2012).


\bibitem{Mannhart} 
 J. Mannhart, and D. G. Schlom,  Science {\bf 327}, 1607 (2010).

\bibitem{jpcm2010} 
 R. Pentcheva, and W. E. Pickett, J. Phys.: Condens. Matter {\bf 32}, 043001 (2010).

\bibitem{zubko} 
  P. Zubko, S. Gariglio, M. Gabay, P. Ghosez, and J.-M. Triscone, Annu. Rev. Condens. Matter Phys. {\bf 2}, 141 (2011).


\bibitem{nickelates} J. Chaloupka, and G. Khaliullin, Phys. Rev. Lett. {\bf 100}, 016404 (2008).


\bibitem{Middey2012}
S. Middey, D. Meyers, M. Kareev, E. J. Moon, B. A. Gray, X. Liu, J. W. Freeland and J. Chakhalian, Appl. Phys. Lett. {\bf 101}, 261602 (2012).

\bibitem{Gibert2012}
 M. Gibert, P. Zubko, R. Scherwitzl, J.-M. Triscone, Nat. Mater. {\bf 11}, 195 (2012).
 
\bibitem{Haldane88}
  F. D. M. Haldane, Phys. Rev. Lett. {\bf 61}, 2015 (1988).
 
\bibitem{KaneMele} 
  C. Kane and E. Mele, Phys. Rev. Lett. {\bf 95}, 226801 (2005).
  
\bibitem{Raghu2008} 
  S. Raghu, X. L.  Qi, C. Honerkamp, and S.-C. Zhang, Phys. Rev. Lett. {\bf 100}, 156401 (2008).

\bibitem{Wright2013} 
  A. R. Wright, 	Sci. Rep.  {\bf 3}, 2736 (2013).

\bibitem{Xiao2011}
    D. Xiao, W. Zhu,	Y. Ran,	N. Nagaosa and S. Okamoto, Nature Commun. {\bf 2}, 596 (2011).

\bibitem{Yang2011}
  K.-Y. Yang, W. Zhu, D. Xiao, S. Okamoto, Z. Wang, and Y. Ran, Phys. Rev. B {\bf 84}, 201104(R) (2011).

\bibitem{Ruegg2011}
   A. R\"uegg, and G. A. Fiete, Phys. Rev. B {\bf 85}, 245131 (2011).

\bibitem{Lado2013}
  J. L. Lado, V. Pardo, and D. Baldomir, Phys. Rev. B {\bf 88}, 155119 (2013).

\bibitem{Doennig2013}   
  D. Doennig, W. E. Pickett and  R. Pentcheva, Phys. Rev. Lett. {\bf 111}, 126804 (2013).

\bibitem{Doennig2014} 
  D. Doennig, W. E. Pickett, and R. Pentcheva, Phys. Rev. B {\bf 89}, 121110 (2014).

\bibitem{Chang2013}
C.-Z. Chang, \etal, Science {\bf 340}, 167 (2013).

\bibitem{Cook2014}
A. M. Cook, A. Paramekanti, Phys. Rev. Lett. {\bf 113}, 077203 (2014).

\bibitem{Garrity2014}
K. F. Garrity and D. Vanderbilt, Phys. Rev. B {\bf 90}, 121103(R) (2014).

\bibitem{Zhang2014}
H. Zhang, J.  Wang, G. Xu, Y. Xu and S.-C. Zhang, Phys. Rev. Lett. {\bf 112}, 096804 (2014). 

\bibitem{wien2k}
  P. Blaha, K. Schwarz, G. K. H. Madsen, D. Kvasnicka, and J. Luitz, 
{\it WIEN2k, An Augmented Plane Wave Plus Local Orbitals Program for Calculating Crystal Properties}, 
   ISBN 3-9501031-1-2
   (Vienna University of Technology, Vienna, Austria, 2001).

\bibitem{pbe96}
 J. P. Perdew, K. Burke and M. Ernzerhof, Phys. Rev. Lett. {\bf 77}, 3865 (1996).

\bibitem{anisimov93} 
V. I. Anisimov, J. Zaanen, and O. K. Andersen, Phys. Rev. B {\bf 44}, 943 (1991).

\bibitem{mBJ}
F. Tran and P. Blaha, Phys. Rev. Lett. {\bf 102}, 226401 2009.

\bibitem{wannier90}
A. A. Mostofi, J. R. Yates, Y. S. Lee, I. Souza, I. Vanderbilt, N. Marzari, Comput. Phys. Commun. {\bf 178}, 685 (2008).

\bibitem{wannierAHC}
X. Wang,  J. R. Yates, I. Souza, I. Vanderbilt, Phys. Rev. B {\bf 74}, 195118 (2006).

\bibitem{wien2wannier} 
J. Kune\v{s}, R. Arita, P. Wissgott, A. Toschi, H. Ikeda, K. Held, Comp. Phys. Commun. {\bf 181}, 1888 (2010).

\bibitem{Torrance1993}
T. Arima, Y. Tokura and J. B. Torrance, Phys. Rev. B {\bf48}, 17006 (1993).

\bibitem{pavarini2004}
 E. Pavarini, S. Biermann, A. Poteryaev, A. I. Lichtenstein, A. Georges, and O. K. Andersen, Phys. Rev. Lett. {\bf 92}, 176403 (2004).
\bibitem{Raychaudhury2007} 
 M. De Raychaudhury, E. Pavarini, O.K. Andersen, 
  Phys. Rev. Lett.  {\bf99},126402  (2007). 

\bibitem{cobalt} H. Hsu, P. Blaha and R. M. Wentzcovitch, Phys. Rev. B {\bf85}, 140404 (2012).

\bibitem{karppinen}
 M. Karppinen, H. Yamauchi, T. Ito, H. Suematsu, O. Fukunaga, Matl. Sci. \& Eng. {\bf 41}, 59 (1996).





\end{thebibliography}
\end{document}